\documentstyle{article}
\begin{document}
\def\bbox {   } 
\def\B.#1{{\bf#1}} 
\title{The Universal Scaling Exponents of
  Anisotropy in Turbulence and their Measurement} 
\author {Victor L'vov and Itamar Procaccia
\thanks{Department of Chemical Physics, The Weizmann Inst. of Science,
 Rehovot 76100, Israel}}
\maketitle
\begin{abstract}
  The scaling properties of correlation functions of non-scalar fields
  (constructed from velocity derivatives) in isotropic hydrodynamic
  turbulence are characterized by a set of universal exponents.  It is
  explained that these exponents also characterize the rate of decay
  of the effects of anisotropic forcing in developed turbulence. This
  set has never been measured in either numerical or laboratory
  experiments. These exponents are important for the general theory of
  turbulence, but also for modeling anisotropic flows. We propose in
  this letter how to measure these exponents using existing data bases
  of direct numerical simulations and by designing new laboratory
  experiments.
\end{abstract}
Fundamental studies of turbulence tend to stress the model of
isotropic, homogeneous turbulence, and most theories and experiments
since Kolmogorov's seminal work of 1941 \cite{41Kol} considered the
universal (anomalous) exponents that characterize the isotropic
characteristics of turbulent flows (see for example \cite{Fri,94Nel}
for recent reviews).  In fact, most turbulent flows are not forced
isotropically, and moreover even in isotropic flows there are
important fields that are constructed from velocity derivatives that
transform under rotation as vectors or tensors rather than scalars.
It has been known for quite a while that the second-order structure
function (that depends on one separation vector) becomes more and more
istotropic as the the separation scale goes down (see below).
Moreover, the rate of this isotropization process is governed by a
universal exponent \cite{67Lum,84NN,94GLLP,95FL}. In recent papers
\cite{95FGLP,96LPP} it was pointed out that this exponent is one of an
infinite family of universal anomalous scaling exponents that were
never considered in experiments and numerical simulations.  Moreover,
it was shown that in the context of passive scalar convection the
anomalous scaling exponents that characterize the scaling properties
of anisotropic fields also govern the rate of isotropization of the
properties of the flow in the cascade process down to smaller and
smaller scales \cite{95FGLP}.  In this Letter we show that the same
connection between the exponents of tensor fields and the exponents
governing the rate of isotropization exists also in Navier-Stokes
turbulence. This demonstration follows from the assumption of ``weak
universality" of hydrodynamic turbulence that is discussed in
\cite{96LP}.  Theoretically the exponents discussed here are related
to the appearance of the inner, viscous scale $\eta$ as a
renormalization scale in addition to the more commonly known
appearance of the outer, integral scale of turbulence $L$
\cite{95LP-2,95LP-3,96LPP}. In addition to being of fundamental
interest these universal properties are also of importance in modeling
realistic flows which are not isotropic.  In this Letter we also
propose how to measure these universal exponents in numerical and in
physical experiments. We will focus here on stationary, space
homogeneous turbulence which is however not necessarily isotropic.

The simplest statistical quantity that is built from the fundamental
velocity field $\B.u(\B.r,t)$ that displays important contributions
from anisotropy is the second-order structure functions of velocity
differences $\B.w(\B.r_0|\B.r,t)\equiv \B.u(\B.r,t)-\B.u(\B.r_0,t)$:
\begin{equation}
S_2(\B.R) \equiv \left < |\B.w(\B.r_0|\B.r,t)|^2\right> \ , \quad
\B.R\equiv \B.r-\B.r_0 \ .\label{S2}
\end{equation}
where $\left<\dots\right>$ stands for a suitably defined ensemble
average.  Due to space and time homogeneity $S_2$ is a time
independent function of the vector $\B.R$.  In isotropic turbulence
the scaling properties of $S_2(R)$ were widely discussed
\cite{Fri,94Nel,75MY}
\begin{equation}
\tilde S_2(R) \simeq (\bar\epsilon R)^{2/3} \left({R \over L}\right)^{\delta}
\ , \label{Sscalar}
\end{equation}
where $\bar\epsilon$ is the mean energy flux per unit time per unit
mass, and $\delta$ is the deviation of the scaling exponent $\zeta_2$
of the structure function from the Kolmogorov 1941 (K41) prediction
$\zeta_2\equiv 2/3-\delta$. In anisotropic turbulence $S_2(\B.R)$
depends on the direction of $\B.R$, and we can decompose it into
spherical harmonics according to the ``multipole expansion"
\begin{eqnarray}
S_2(\B.R)  &=&
\sum_{l=0}^\infty  S_{2,\ell} ( \B.R) \ ,  \label{multipole}\\
S_{2,\ell} ( \B.R) &=& \sum_{m=-l}^l Y_{lm} ( \hat\B.R)
\int S_2(R\hat{\bbox {\xi}})
 Y_{lm} ( \hat {\bbox {\xi}}) d \hat{\bbox {\xi}}  \  , \label{psil}
\end{eqnarray}
where $\hat{\bbox{\xi}}$ is a unit vector. In a scale invariant situation
every component
$S_{2,\ell}$  scales like
\begin{equation}
S_{2,\ell}(\B.R)\sim \left (\bar\epsilon R\right )^{2/3} \left({R\over
L}\right)^{\delta_{\ell}}
\propto R^{\beta_{\ell}}\ , \label{scale}
\end{equation}
where $\delta_{\ell}\equiv \beta_{\ell}-2/3$. Comparing with
Eq.(\ref{Sscalar}) we recognize that in this notation
$\delta=\delta_0$. The full spectrum of exponents $\beta_{\ell}$ was
found analytically \cite{95FGLP,95CFKL} in the context of Kraichnan's
model of passive scalar convection\cite{94Kra}. For Navier-Stokes
turbulence $\beta_2$ can be computed using perturbation theory
\cite{94GLLP,95FL} (which disregards the nonperturbative effects
leading to anomalous scaling \cite{95LP-3}) with the result $\beta_2=
4/3$.  The corresponding result for $\beta_0\equiv \zeta_2$ is 2/3,
which experimentally is known to be of the order of 0.7
\cite{87MS,93BCTBM}. The theory indicates that such deviations from
the naive predictions stem from non-perturbative effects. It is likely
that the perturbative result for the exponent $\beta_2$ holds to a
similar accuracy. The large difference between $\beta_2$ and $\beta_0$
explains why isotropic scaling may be observed in anisotropic
experiments; the contribution of $S_{2,2}$ to $S_2$ peels off rather
quickly when $R\ll L$.  We do not possess any numerical estimates for
the higher order values of $\beta_{\ell}$, but we expect them to be
all positive and increasing with $\ell$.

In the context of passive scalar convection we demonstated that the
very same exponents $\beta_{\ell}$ have an important role in the
context of {\em isotropic} turbulence when we considered statistical
quantities that depend on more than two coordinates\cite{95FGLP}.  We
point out here that the same is true for Navier-Stokes turbulence.
Consider for example the correlation function of four velocity
differences
\begin{equation}
{  S}_4(\B.R_1,\B.R_2) \equiv
\left<|{\bf w}({\bf r}_0|{\bf r}_1)|^2|{\bf w}({\bf r}_0|{\bf
r}_2)|^2\right> \ , \label{S4}
\end{equation}
where $\B.R_1=\B.r_1-\B.r_0$ and $\B.R_2=\B.r_2-\B.r_0$.  As usual we
assume that this, and all other correlators, are scale invariant.
Mathematically this means that they are all homogeneous functions of
their arguments as long as these are in the ``inertial range". In
other words ${ S}_4(\lambda\B.R_1,\lambda\B.R_2) =\lambda^{\zeta_4} {
  S}_4(\B.R_1,\B.R_2)$ where $\zeta_4$ is the scaling exponents of the
4'th order structure function: $\left<|{\bf w}({\bf r}_0|{\bf
    r}_1)|^4\right>\propto R_1^{\zeta_4}$.  In isotropic turbulence ${
  S}_4(\B.R_1,\B.R_2)$ depends on the separations $R_1, R_2$ and on
the angle $\theta_{1,2}$ between these two vectors. We are interested
in the limit $R_1\ll R_2$, but $R_1$ and $R_2$ are both in the
inertial interval.  It was shown in \cite{95LP-3,96LP} that in this
limit the leading dependence of ${ S}_4(\B.R_1,\B.R_2) $ on $R_1$ and
$R_2$ is independent of $\theta_{1,2}$ and that it scales like
\begin{equation}
{  S}_4(\B.R_1,\B.R_2) \propto \left({R_1\over
R_2}\right)^{\zeta_2}R_2^{\zeta_4} \ ,
\quad R_1\ll R_2\ . \label{fuse}
\end{equation}
In order the extract the sub-leading dependence on $\B.R_1$ we use a
mulitpole decomposition of ${ S}_4$ in a way similar to (\ref{psil}):
\begin{eqnarray}
{  S}_4(\B.R_1,\B.R_2)  &=&
\sum_{l=0}^\infty  S_{4,\ell} ( \B.R_1,\B.R_2) \ ,  \label{mult2}\\
{  S}_{4,\ell} ( \B.R_1,\B.R_2) &=& \sum_{m=-l}^l Y_{lm} ( \hat\B.R_1)
\int S_2(R_1\hat{\bbox {\xi}},\B.R_2)
 Y_{lm} ( \hat {\bbox {\xi}}) d \hat{\bbox {\xi}}  \  . \nonumber
\end{eqnarray}
The generalization of (\ref{fuse}) to the anisotropic contributions is
\begin{equation}
{  S}_{4,\ell}(\B.R_1,\B.R_2) \propto \left({R_1\over
R_2}\right)^{\beta_{\ell}}R_2^{\zeta_4},
\quad R_1\ll R_2\ .\label{fusean}
\end{equation}
Note that there are two statements made here. The first is that the
over-all exponent for this quantity is $\zeta_4$. This directly
follows from the property of scale invariance.  The second statement
is that the scaling exponents characterizing the $R_1$-dependence of
${ S}_{4,\ell}$ are the scaling exponents of ${ S}_{2,\ell}$.  This
result follows from the assumption of ``weak universality" used in the
derivation of the fusion rules in \cite{96LP}. Physically it is
equivalent to the statement that the measured scaling exponents of the
structure functions in turbulence are independent of the precise
driving mecahnism at the outer scale of turbulence $L$. The existence
of $R_2$-scale eddies and their effect on the statistics of the much
smaller $R_1$ eddies is similar to the existence of $L$-scale eddies
and their effect on the scaling exponents in the inertial range. The
velocity difference measured at points $\B.r_1$ and $\B.r_0$ is
effected by the large eddies characterizing the velocity difference
across $\B.R_2$ in a way that is similar to the effect of the boundary
conditions at $L$ on the velocity difference measured in $S_2$.
Similarly the exponents $\beta_{\ell}$ characterize the rate of
isotropization of $S_4$ as a function of $\B.R_1$. We stress that this
isotropization is relevant in isotropic turbulence since we have in
this function a built-in direction $\B.R_2$.  When $R_1$ is of the
order of $R_2$ the dependence on the angle $\theta_{1,2}$ is all
important. When $R_1$ decreases this dependence weakens at a rated
determined by the exponents $\beta_{\ell}-\beta_0$.

Next in order of complication we consider $S_4(\B.R_1,\B.R_2,\B.R_0)$
defined as
\begin{equation}
{  S}_4(\B.R_1,\B.R_2,\B.R_0) \equiv
\left<|{\bf w}({\bf r}_0|{\bf r}_1)|^2|{\bf w}({\bf r}'_0|{\bf
r}_2)|^2\right> \ , \label{S4RRR}
\end{equation}
where $\B.R_0=\B.r'_0-\B.r_0$.  This is a function of three separation
and the three angles $\theta_{1,0}$, $\theta_{2,0}$ and
$\theta_{1,2}$.  As before represent this function as a double
multipole-expansion with respect to the directions of $\B.R_1$ and
$\B.R_2$:
\begin{equation}
{  S}_4(\B.R_1,\B.R_2,\B.R_0) = \sum_{\ell_1,\ell_2}
S_{4,\ell_1,\ell_2}(\B.R_1,\B.R_2,\B.R_0)
\ . \label{doubmul}
\end{equation}
In the limit $R_1,R_2 \ll R_0$ these functions exhibit a universal
scaling form similar to (\ref{fusean})
\begin{equation}
S_{4,\ell_1,\ell_2}(\B.R_1,\B.R_2,\B.R_0)\propto
\left({R_1\over R_0}\right)^{\beta_{\ell_1}}\left({R_2\over R_0}\right)^
{\beta_{\ell_2}}R_0^{\zeta_4} \ .
\label{fuse2}
\end{equation}

Finally we discuss correlations of anisotropic local fields
constructed from velocity derivatives.  These can be obtained by a
limiting procedure starting from the fusion of two points in the
vicinity of $\B.r_0$ as in (\ref{fusean}) or in the vicinity of
$\B.r_0$ and $\B.r'_0$ as in (\ref{fuse2}). As shown in \cite{96LPP}
the simplest representatives of such fields contain a product of two
derivatives $\partial_\alpha u_\beta \partial_\gamma u_\delta$. Higher
tensorial fields are obtained by taking additional derivatives from
this field. To get clean scaling behaviour we need to decompose these
fields to combinations that give irreducible representations of the
rotations and inversion group O(3). Every irreducible representation
is characterized by an index $\ell$, has a dimension $(2\ell+1)$, and
the $(2\ell+1)$ fields that form its basis transform like the
spherical harmonic $Y_{\ell,m}$. The low orders representations are
constructed with the help of the strain tensor
$s_{\alpha\beta}=[\partial u_\alpha / \partial r_\beta+\partial
u_\beta / \partial r_\alpha]/2$ and the vorticity field
$\B.\omega_\alpha=\epsilon_{\alpha\beta\gamma} \partial u_\beta /
\partial r_\gamma$ (where $\epsilon_{\alpha\beta\gamma}$ is the fully
antisymmetric pseudo-tensor). There are two scalar fields,
$O^{(1)}_0\equiv\omega_\alpha\omega_\alpha$ and $O^{(2)}_0\equiv
s^2=s_{\alpha\beta}s_{\beta\alpha}$ each of which is a basis for
one-dimensional irreducible representation with $\ell=0$.  The
pseudo-vector $O^\alpha_1\equiv s_{\alpha\beta}\omega_\beta$ is a
three-dimensional basis for an irreducible representation with
$\ell=1$. There exist three traceless tensor fields each of which is a
five-dimensional basis belonging to $\ell=2$ and taking care of
$3\!\times\!5=15$ components. An example is
\begin{equation}
O_2^{(1)\alpha\beta}(\B.r)=s_{\alpha\gamma}(\B.r)
s_{\gamma\beta}(\B.r)-\delta_{\alpha\beta}
s^2(\B.r)/3 \ . \label{O2}
\end{equation}
In addition we have one 3-rank pseudo tensor corresponding to $\ell=3$
and one 4-rank tensor corresponding to $\ell=4$. The last two fields
exhaust the remaining $7+9$ components \cite{96LPP}.  We note that all
the tensor fields $\B.O_\ell$ are dimensionally identical. However,
dimensional analysis misses the point, and fields that transform
differently under the symmetry group have different scaling exponents.
The correlation functions $\left < \B.O_\ell(\B.r+\B.R)
  \B.O_{\ell'}(\B.r)\right >$ all have different scaling exponents
depending on $\ell$ and $\ell'$:
\begin{equation}
\B.K_{\ell\ell'}(\B.R)\equiv \left < \B.O^{(n)}_\ell(\B.r+\B.R)
\B.O^{(n')}_{\ell'}(\B.r)\right >
\propto R^{-\mu_{\ell\ell'}} \ ,\label{OO}
\end{equation}
independent of $n$ and $n'$. In particular the prediction is that the
three correlation functions involving the scalar fields $s^2$ and
$\omega^2$ have the same scaling exponents known as the
``intermittency exponent" $\mu$.  Similarly the six correlations
involving $\B.O_2$ have the same exponent (different from $\mu$).
Note that the rank of $\B.K_{\ell\ell'}$ is the sum of the ranks of
the tensors in the correlation.

At this point we want to discuss how to set up possible experiments to
measure the new universal exponents $\beta_{\ell}$. Given a direct
numerical simulation with anisotropic forcing, the most
straightforward way it to simply compute $S_{2,\ell}(\B.R)$ from the
definitions (\ref{S2}) and (\ref{psil}) and then to plot log-log plots
of $S_{2,\ell}$ vs. $R$, or even better, following the ideas
\cite{93BCTBM} of ``extended self-similarity", of $S_{2,\ell}$ vs.
$S_{2,\ell=0}$. It is impossible to follow this route in standard
laboratory experiments since the detailed angular information is not
usually available. One can estimate the exponent $\beta_2$ in
anisotropic flows by measuring for example the longitudinal and
transverse components of the second order structure function, and form
a combination that vanishes in isotropic flows. Such a combination
scales with $R$ and the leading contribution is $R^{\beta_2}$.  This
type of measurement was performed, see for example \cite{94SV} and
discussed in detail by Nelkin \cite{94Nel}. The experimental evidence
is that the numerical value of $\beta_2$ is indeed rather close to
4/3. Our point in this Letter is that the very same exponents
$\beta_\ell$ play an important role also in isotropic flows.

In laboratory (and also in atmospheric) experiments it is difficult to
resolve the dissipative scales, and the direct measurements of the
local fields $O_{\ell}(\B.r)$ is quite hard.  Accordingly we will
suggest experiments that are based on finite differences in the
inertial range instead of gradients.  Consider an experiment with a
mean flow (like a wind tunnel or an atmospheric boundary layer).
Assign the direction of the mean flow to the $x$-coordinate.  The
minimal experimental set up calls for two local probes (like hot
wires) positioned at $\B.r_0=(0,0,0)$ and $\B.r_1=(0,\Delta,0)$,
separated by a distance $\Delta$ in the $y$-direction which is
orthogonal to the mean flow. Under the standard Taylor hypothesis
differences in time are interpreted as differences along the
longitudinal $x$-direction. This means that one can measure the
longitudinal projections $a u_x(x,0,0)$ and $b u_x(x,\Delta,0)$. The
coefficients $a$ and $b$ were introduced in recognition of the fact
that in realistic experiments the two probes cannot be perfectly
calibrated.  Define now the longitudinal and transverse velocity
differences
\begin{eqnarray}
w_{||}(x,\Delta)&\equiv& u_x(x+\Delta,0,0)-u_x(x,0,0)\ , \\
w_{\perp}(x,\Delta) &\equiv& u_x(x,\Delta,0)- u_x(x,0,0)\ . \label{lon}
\end{eqnarray}
Next one can measure the longitudinal and
transverse structure functions for $\Delta$-separations
\begin{equation}
S_{2||}(\Delta)\equiv \langle w^2_{||}(x,\Delta) \rangle,\quad
S_{2\perp}(\Delta)\equiv
\langle w^2_{\perp}(x,\Delta) \rangle \ . \label{Slon}
\end{equation}
In isotropic conditions these two quantities are related \cite{75MY}
by
\begin{equation}
S_{2\perp}(\Delta) = S_{2||}(\Delta)+\Delta dS_{2||}(\Delta)/2d\Delta \ ,
\label{relat}
\end{equation}
and one can use this relation to assess the degree of isotropy on the
scale $\Delta$.  Next we introduce the normalized squared of velocity
differences in which the calibration constants are eliminated:
\begin{equation}
W^2_{||} (x,\Delta)\equiv {w^2_{||}(x,\Delta)\over S_{2||}(\Delta)}\ , \quad
W^2_{\perp} (x,\Delta)\equiv {w^2_{||}(x,\Delta)\over S_{2\perp}(\Delta)} \ .
\label{W2}
\end{equation}
Finally we define two fields $\Psi_{\pm}$ according to
\begin{eqnarray}
\Psi_{+}(x,\Delta)&\equiv& W^2_{\perp} (x,\Delta)+W^2_{||} (x,\Delta)-2\ ,
\nonumber\\
\Psi_{-}(x,\Delta)&\equiv& W^2_{\perp} (x,\Delta)-W^2_{||} (x,\Delta) \ .
\label{psi}
\end{eqnarray}
These fields have zero mean by construction, and we propose to measure
their correlations $K_{++}(R)$, $K_{--}(R)$ and $K_{+-}(R)$ across a
scale $R$ such that $\eta <\Delta <R <L$.  The theoretical prediction
is that the leading scaling form is
\begin{eqnarray}
K_{++}(R)&\equiv& \left <\Psi_{+}(x+R,\Delta) 
\Psi_{+}(x,\Delta) \right> \propto
R^{-\mu_{++}}\ ,\label{++}\\
K_{+-}(R)&\equiv& \left <\Psi_{+}(x+R,\Delta) 
\Psi_{-}(x,\Delta) \right> \propto
R^{-\mu_{+-}}\ ,\label{+-}\\
K_{--}(R)&\equiv& \left <\Psi_{-}(x+R,\Delta) 
\Psi_{-}(x,\Delta) \right> \propto
 R^{-\mu_{--}}\ , \label{--}
\end{eqnarray}
where the exponents are
\begin{eqnarray}
\mu_{++}&=&2\beta_0-\zeta_4 \ , \quad \beta_0\equiv \zeta_2\ , 
\nonumber  \\
\mu_{+-}&=&\beta_0+\beta_2-\zeta_4\ , \quad
\mu_{--}=2\beta_2-\zeta_4 \ .
\end{eqnarray}
The reason for this prediction is that the field $\Psi_{+}$ has a
large projection on the zero'th spherical harmonic $Y_{0,m}$, whereas
the field $\Psi_-$ has no such projection by construction. Moreover
the field $\Psi_-$ also has no projection on $Y_{1,m}$. It does have
projections on $Y_{2,m}$ and higher order spherical harmonics. Thus
the leading scaling exponents appearing in correlations of $\Psi_-$ is
$\beta_2$, whereas the leading scaling exponent appearing in
correlations of $\Psi_+$ is $\beta_0$, and Eqs.(\ref{++})-(\ref{--})
follow directly from (\ref{fuse2}). Note that the correlations in
(\ref{++})-(\ref{--}) are all dimensionally identical; yet we predict
very different scaling exponents.  This is just another way to explore
the breakdown of dimensional analysis in fully developed turbulence.
Using known experimental data \cite{87MS,93BCTBM} $\zeta_2\simeq 0.7$,
$\zeta_4\simeq 1.2$ and our guess that $\beta_2\simeq 4/3$ we expect:
\begin{equation}
\mu_{++}\simeq 0.2,~~ \mu_{--}\simeq 1.4-1.5 , ~~
\mu_{+-}={\mu_{++}+\mu_{--}\over 2}
\ . \label{mus}
\end{equation}
The last relation is asymptotically (in Re) exact. Note however that
for a finite extent of the inertial interval sub-leading scaling
contributions may be important and have to be carefully assessed.
Nevertheless, the wide disparity between these scaling exponents
promises a worthwhile experiment even if the inertial range is of the
order of one decade.

Direct numerical simulations offer additional ways to examine the
correlation function $\B.K_{\ell\ell'}(\B.R)$ (\ref{OO}) of the
anisotropic local fields.  It was pointed to us by R. Benzi \cite{Ben}
that in numerical simulations the correlation function of gradient
fields have a scaling range that is too short.  Indeed, even the
correlation function of the scalar dissipation $\epsilon(\B.r)=2\nu
s^2(\B.r)$ is not readily available.  It was suggested that a better
scaling behaviour is exhibited by integrals of the dissipation field
over balls of radius $R$. The calculation is achieved by first
considering
\begin{equation}
 \epsilon_R(\B.r_0) = {3\over 4\pi R^3}\int_{|\B.r-\B.r_0|\le
R}d\B.r~\epsilon(\B.r) \ ,
\label{eR}
\end{equation}
from which one computes the time and space average
\begin{equation}
\left< \hat\epsilon^2_R \right > \equiv \left<
\left[\epsilon_R(\B.r_0)-\bar\epsilon\right]^2\right>
\ . \label{e2R}
\end{equation}
For $R$ in the inertial interval
\begin{equation}
\left< \hat\epsilon^2_R \right > = C_\mu K_{\epsilon\epsilon}(R)\propto
R^{-\mu} \ ,
\label{bla}
\end{equation}
where $C_\mu \simeq 1$ and
$K_{\epsilon\epsilon}(R)=\left<[\epsilon(\B.r_0+\B.R)-\bar\epsilon][\epsilon
  (\B.r_0)-\bar\epsilon] \right> $.  Following the same idea
\cite{Ben} instead of computing in simulations the correlations
$\B.K_{\ell\ell'} (\B.R)$ we can measure the mean value of the product
of two $R$-ball integrated fields
\begin{equation}
 \B.O_{\ell,R}(\B.r_0) = {3\over 4\pi R^3}\int_{|\B.r-\B.r_0|\le
R}d\B.r~\B.O_{\ell}(\B.r) \ . \label{OlR}
\end{equation}
Accordingly
\begin{equation}
\left< \B.O_{\ell,R}(\B.r_0) \B.O_{\ell',R}(\B.r_0)\right> = C_{\ell\ell'}
\B.K_{\ell\ell'}(\B.R)\propto
R^{-\mu_{\ell\ell'}} \ .
\label{wow}
\end{equation}
It should be stressed here that we did not offer yet any numerical
estimates for $\mu_{\ell\ell'}$. The reason is deep: the local fields
pick up a dissipative scale when the gradient is computed. In
Navier-Stokes turbulence there is a multiplicity of dissipative scales
that have non-trivial dependencies on the inertial-range separation
distances which the relevant correlation function depends on. The full
discussion of this issue will be available elsewhere \cite{96LP-4}.
Here we will just present theoretical prediction without derivation.
They are:
\begin{equation}
\mu_{\ell\ell'} =2-\zeta_6+\beta_\ell+\beta_{\ell'}-2\beta_0 \ . \label{pred}
\end{equation}
For $\ell=\ell'=0$ we recover the well known ``bridge relation" which
followed from the Kolmogorov refined similarity hypothesis,
$\mu=2-\zeta_6$.  The predictions for non-zero $\ell$ are novel and
await confirmation. We stress that such a measurement can give
information about $\beta_1$ which is associated with the pseudo-vector
field $s_{\alpha\beta}\omega_\beta$. This exponent is not available
from the rate of isotropization of $S_2(\B.R)$. It can be seen if the
flow field does not have inversion symmetry but one needs to form a
nonsymmetric second-order correlation function like
\begin{equation}
\left<u_\alpha(\B.r+\B.R)u_\beta(\B.r)\right
>-\left<u_\alpha(\B.r-\B.R)u_\beta(\B.r)\right >\propto
R^{\beta_1} \ .
\end{equation}
Since this object is manifestly odd in $\B.R$ it vanishes when there
exists inversion symmetry.  Otherwise its leading scaling exponent is
$\beta_1$. This exponent is related to the existence of the flux of
helicity and standard K41 arguments lead to the value $\beta_1=1$, see
for example \cite{81KL}. This holds probably to the same accuracy as
other K41 arguments. The correlation functions discussed above allow
the measurement of this exponent in the presence of inversion
symmetry. Even thought $\B.K_{\ell\ell'}$ will vanish for $\ell$ even
and $\ell'$ odd, the correlation $\B.K_{11}$ is non-zero in any case.

In summary we presented briefly the ideas that related the infinite
set of universal exponents characterizing the rate of isotropization
of the simple second-order structure function under non-isotropic
forcing with the same set of exponents that determines the scaling
behaviour of correlation functions of tensorial anisotropic local
fields in isotropic turbulence. The central role that these exponents
play warrants their measurement in laboratory and numerical
experiments. We thus offered some simple ways to measure the low order
exponents $\beta_1$ and $\beta_2$ in realistic experiments. We
presented an estimate of the numerical values of these two exponents.
The calculation of these exponents from first principles is a
different task that is outside the scope of this Letter.
\vskip.5cm

{\bf Acknowledgments} 
This paper was motivated by discussions with Roberto
Benzi, to whom we offer thanks. We profitted from discussion with
Evgeni Podivilov and Victor Steinberg. This work has been supported in
part by the US-Israel Bi-National Science Foundation, the
German-Israeli Foundation and the Naftali and Anna Backenroth-Bronicki
Fund for Research in Chaos and Complexity.

\end{document}